\documentclass{aa} 

\usepackage{graphicx}

\usepackage{longtable,lscape} 

\usepackage{lscape}

\usepackage{txfonts}

\usepackage{color}

\begin{document}

\title{A two-mode planetary nebula luminosity function} 
\author{A. Rodr\'\i guez-Gonz\'alez\inst{1},
L. Hern\'andez-Mart\'inez\inst{2,3}, A. Esquivel\inst{1},
A.C. Raga\inst{1}, G. Stasi\'nska\inst{4}, M. Pe\~na\inst{3} \& D. 
Mayya\inst{2}} 

\offprints{A. Rodr\'iguez-Gonz\'alez} 

\institute{
1 Instituto de Ciencias Nucleares, Universidad Nacional Aut\'onoma 
de M\'exico, Ap. 70-543, 04510 D.F., M\'exico.\\
2 Instituto de Astrof\'isica, \'Optica y Electr\'onica, Ap. 51, 72000 
Puebla, M\'exico.\\
3 Instituto de Astronom\'\i a, Universidad Nacional Aut\'onoma de M\'exico, 
Ap. 70-264, 04510  D.F., M\'exico.\\ 
4 LUTH, Observatoire de Meudon, 92195, Meudon Cedex, France\\
\email{ary@nucleares.unam.mx} } 
\date{Received xxxxxxxxxxx; accepted xxxxxxxxxxxx} 

\titlerunning{Two mode PNLF}

\authorrunning{Rodr\'iguez-Gonz\'alez et al.} 


\abstract 
{We propose a  new Planetary Nebula Luminosity Function (PNLF) that
  includes two populations in the distribution. Our PNLF is a
  direct extension of the  canonical function proposed by Jacoby et
  al. (1987), in order to avoid problems related with the histogram
  construction, it is cast in terms of cumulative functions.}
{We are interested in recovering the shape of the faint part of the
  PNLF in a consistent manner, for galaxies with and without a dip in
  their PN luminosity functions.}
{The parameters for the two mode PNLF are  obtained with a genetic
  algorithm, which obtains a best fit to the PNLF varying all of the
  parameters simultaneously in a broad parameter space.}
{We explore a sample of 9 galaxies with various Hubble types and
  construct their PNLF. All of the irregular galaxies, except one, are
  found to be consistent with a two-mode population, while the
  situation is less clear for ellipticals and spirals.For the case of
  NGC\, 6822, we show that the two-mode PNLF is consistent with
  previous studies of the star formation history within that
  galaxy. Our results support two episodes of star formation, in which
  the latter is significantly stronger.}
{}

\keywords{ ISM: planetary nebulae: general --galaxies: luminosity function, mass function} 

\maketitle


\section{Introduction}
More than two decades ago, several authors reported
 that the bright end of the distribution function of
planetary nebula (PNe), as a function of their magnitude in the
[O\,{\sc iii}] 5007 line, was remarkably similar in  a large sample of
galaxies (Jacoby et al. 1990a, 1990b; Ciardullo 2012, and reference therein). 
The galaxies studied were mainly ellipticals, irregulars and a few spirals.
In particular, they found that the bright cutoff of the planetary
nebula luminosity function (PNLF)  is
consistent with a universal absolute magnitude $M_{5007}^*$, from which
a distance modulus can be calculated, and thus PNLFs have been used
as standard candles to estimate the distance to galaxies where PNe are
observed.

Among such studies are  those by  Jacoby et
al. (1989) for M\,81, Ciardullo et al.  (1989b) for the Leo I group,
Jacoby et al. (1990a) for the Virgo Cluster, Pottasch (1990) for the
galactic center, Jacoby et al. (1990b) for the Magellanic Clouds,
Ciardullo et al. (1991) for the NGC\,1023 group, McMillan et al. (1993)
for the Fornax Cluster, Jacoby et al. (1996) for the Coma I region,
Feldmeier al. (1997) for several spiral galaxies, Pe\~na et al. (2007
for NGC\,3109, Herrmann et al. (2008) for several face-on spiral
galaxies, Hern\'andez-Mart\'inez \& Pe\~na (2009) for NGC\,6822, Pe\~na et
al. (2012) for NGC\,300, and several others. 

A number of studies attempted to understand how, why,
or in what type of galaxies the PNLF cutoff can indeed be used as a
standard candle for distance estimates (Dopita et al. 1992;  M\'endez
et al. 1993; Ciardullo et al. 1989a and 1989b), however,   the
physical mechanism responsible for such universality in the
M$^*_{5007}$, especially across galaxies of different Hubble types, is
still not clear. 

Theoretical attempts to model the PNLF of an old stellar population,
corresponding to what we expect to find in elliptical galaxies,
assuming single-star post-AGB evolution, have not been successful
(Marigo et al. 2004; Ciardullo 2006). A possible explanation could
involve massive central stars in old populations produced through
binary evolution (e.g. Ciardullo et al. 2005). More recently,
using fully hydrodynamical simulations of the evolution of PNe,
Sch\"onberner et al. (2007) argue that there is no need to invoke  
central stars with masses over 0.7 M$\odot$ to account for the bright
end of the PNLF.  At present the invariance of M$^*_{5007}$ in all
Hubble type galaxies is still  an open question.
 
Henize \& Westerlund (1963) were the first to compute a PN luminosity
function. They assumed that PNe are objects of constant mass and are
subject to  uniform expansion velocity, so that they  fade as a result
of expansion. The number of PNe between magnitudes $M$ and $M+dM$ is
proportional to the time spent in that magnitude range.  Jacoby (1980)
showed that observations of the faint PNe in the LMC fit well  the
luminosity  function proposed by HW63. In general, the faint part of
the observational PNLF follows the  HW63 model, while the brightest
part  decreases  more steeply.

On the other hand, Ciardullo et al. (1989a hereafter C89) showed that
the brightest part  of the PNLF can be reproduced by an exponential cutoff. 
C89, using  the magnitudes of [O\,{\sc iii}] 5007, found a good agreement 
between their theoretical PNLF and observations of PNe in 24 spiral and elliptical galaxies. 
Fitting the exponential cutoff in combination with the HW63 luminosity
function,  C89 suggested that PNe could be used as standard candles. 

While the upper end of the PNLF is invariant among galaxies, its
global shape may vary from one galaxy to another. Some galaxies, like
the Small Magellanic Cloud (SMC, Jacoby \& de Marco 2002, hereafter
JM02)  present a dip in the PNLF, others do not. The details of the
interpretation of this dip vary among authors, but in essence it is
considered to be a result of the rapid decline in luminosity of the
most massive central stars descending the white dwarf cooling track
not being compensated by the presence of PNe with less massive central
stars (Jacoby \& de Marco 2002, Marigo et al 2004, M\'endez et
al. 2008). Thus, the presence or absence of a dip in the PNLF can be a
signature of the star formation history of those stellar populations
that give rise to the PNe detected at present in the galaxies. However, since the canonical PNLF does not account for the presence of
such a dip, it is customary to restrict the data used to derive the
PNLF parameters to the brightest end when it shows evidence of a
decrement in number of PNe towards larger magnitudes.

Hern\'andez-Mart\'inez \& Pe\~na (2009, hereafther HMP09) constructed
the PNLF of the dwarf irregular NGC\,6822  and found  a
statistically significant dip in the luminosity function. In both the  SMC and NGC\,6822 
the PNLF dip is $\sim$2.5~mag fainter than the brightest PNe. 
Carigi et al. (2006) and Hern\'andez-Mart\'inez et al. (2009)
presented a study of the star formation history of NGC\,6822,
showing evidence of two important  star forming episodes, which could be 
related to a dip seen in the PNLF (see section 6).
 A similar dip is seen in the PNLF of other galaxies which are
known to have more than one star formation episode (the SMC for
instance, see N\"oel et al. 2008).

 If the PNe sample is large, obtaining the LF via traditional $\chi^2$ approach
 is relatively simple
because one could fit the PNLF to a histogram built with the data
(i.e. M\'endez et al. 2001, Teodorescu et al. 2005, Johnson  et
al. 2009). 
However, if the sample of PNe is small, this procedure becomes very
sensitive to the parameters used to construct the histogram  (e.g. bin
size, number of bins, range of values used). For instance the presence
of a dip can be missed by a slight change in bin size or position.
In order to minimize this uncertainties C89,  based on Hanes \&
Whittaker 1987, perform a maximum likelihood analysis, from which they
obtain the fitting parameters of the function.
Alternatively, Pe\~na et al. (2007) fitted small samples of PNe to a
\textit{cumulative} luminosity function. 

The cumulative luminosity function is also insensitive to the
histogram parameters, but some important features of the
canonical PNLF could be masked (for instance the dip).

 Another issue with the PNLF is that the sample can be incomplete
at the faint end. In fact, since the most important parameter used
in the canonical PLNF is the magnitude of the brightest PN (see the
following section), usually only the brightest PNe in
the analysis. However, it is not clear what is the appropriate range,
and by restricting the sample one could misidentify a dip with a
decline in number due incompleteness.

We built a cumulative PNLF in order to use small or large samples.
We only restrict the data at the faint end when incompleteness of the
sample is obvious, even if there is evidence for a dip in the sample.

An important goal of this paper is to recover the shape of the faint part of the PNLF in a
consistent manner, for galaxies with and without a dip in their PN
luminosity functions. The proposed PNLF is based on the function by
Ciardullo et al. (1989a). 

The paper is organized as follows: 
In Section~2 we
review the properties of the canonical PNLF and its cumulative form.
A two-mode planetary nebula luminosity function is described in
Section~3, and in Section~4 we present the results of the PNLF for a
sample of 9 galaxies. In section 5 we show that the two modes function can be
consistent with other observational data.
A summary is provided in Section~6.

\section{Cumulative and non-cumulative PNLFs}

Usually larger samples allow for smaller bins to form a
histogram with more detail without loosing its general shape (see
Hogg 2008).
Some studies for large samples suggest that the bin-size should be
related to the dispersion and the total number of data points (Scott 1979).
For instance, for normally distributed samples it has been proposed 
that a bin size
\begin{equation}
  \label{scott}
  \Delta m \propto \frac{\sigma}{n^{1/3}_{data}},
\end{equation}
is appropiate, where $\sigma$ is the standard deviation and $n_{data}$ is the number
of data points in the sample.\\

For the PNLF this is not the case, the optimal bin size to present it
as a histogram can be chosen after the data has been fit, for instance
by a maximum likelihood analysis (see C89). However, it does not
correlate simply with the sample size.

More recently, Herrmann et al. (2008) presented the PNLFs for several
spiral galaxies. All of the observational PNLFs  constructed therein  used 
a uniform bin-size ($\sim 0.3$~mag),  regardless  of the  number  of
PNe in each galaxy  (ranging from 20 to 150 objects). Reid \&
Parker (2010) added 80 PNe to the luminosity function of the LMC,
with 584 objects and a bin-size of 0.2~mag. 

Notice that the bin size is not the only parameter that determines a
histogram, also the maximum and minimum magnitude, as well as the
center of the bins is important.
The selection of the position of the bin center can also be important
to find a dip in the PNLF. If a histogram is used to restrict the data
to be fit, the presence of a dip can be confused with sample
incompleteness and indirectly affect the fit. 

One way to avoid potential binning issues (which has been adopted in
some  studies) is  to use cumulative distribution functions.
For instance, the cumulative function of the PNLF was recently
fitted to a sample of $\sim$20 PNe in NGC\,3109 by Pe\~na et al
(2007). From their estimate of $m^*_{5007}$, the authors obtained a
distance modulus from the  PNLF  in  good agreement with the value
obtained from Cepheid stars.
As we will show in the next section, the fitting of the observational data with
cumulative functions is not affected by the histogram considerations,
such as the bin-size, limits, or the position of the first bin.s

The canonical form of the PNLF (based on the function of HW63, with the
exponential cutoff suggested by C89) is
\begin{equation}
\label{eq:pnlf}
N(m_{5007})=n\,e^{-0.307\mu}e^{0.307m_{5007}}\left[1-e^{3(m^*_{5007}-m_{5007})}\right],
\end{equation}
where $n$ is a normalization constant, $m^*_{5007}$ is the apparent 
$\lambda$5007 magnitude of the brightest  PN  which can exist in
a given galaxy, $\mu=5 \log d -5 + A_{5007}$ is the  distance
modulus, and  A$_{5007}$ is the extinction. In order  to simplify the notation we will substitute,
$N_T=n\, e^{-0.307\mu}$, in  equation~\ref{eq:pnlf} and drop the
$5007$ sub-indices to obtain
\begin{equation}
\label{eq:pnlf2}
N(m;N_T,m^*)=N_T e^{0.307m}\left[1-e^{3(m^*-m)}\right].
\end{equation}

As discussed above,  empirical evidence suggests that the absolute  magnitude $M^*_{5007}$, is the same for all galaxies. 
Ciardullo et al. (2002) derived different M$^*_{5007}$ for different metallicities, the difference is not very significant so PNe 
have been regarded as standard candles and are widely
used in the so called cosmic ladder as a distance indicator.

Several observations of the PNLF in nearby
galaxies (IC\,342, M\,74, M\,83, M\,94, see also Herrmann et al. 2008), in which the canonical luminosity function of Jacoby 
et  al. (1990) is in good agreement with the observed one at least in the brightest 1.5~mag, support the use of this portion of 
the PNLF as a distance indicator.  However, the faintest part  cannot always be reproduced by the exponential shape, and 
almost such all studies blame it to the incompleteness of the sample.  

As we have mentioned  before, the observed  PNLFs  for two irregular galaxies, i.e. SMC, NGC\,6822 
(see JM02, HMP09), show  a ``dip'' at about 2.5~mag after which the PNLF  rises  again and then drops at
the end of the sample. This decrease in the number of
planetary nebulae  could be considered as evidence of an additional
stellar population or a different evolutionary scenario for the central stars in the PNe.
 In these cases, the authors  have also
focused only on the brightest PNe observed to estimate distances.

One of the techniques used to estimate the PNLF is to build a histogram of 
the apparent magnitudes
of the PNe and fit the brightest portion of this histogram to the
functional form of the PNLF to estimate $m^*$ and $N_T$. This procedure 
is particularly tricky because the number of bins, the bin size and the 
initial position of the first bin are treated as free parameters, and they 
are commonly determined arbitrarily. In fact, the choice of bin size and 
position of the first bin can determine whether a dip in the PNLF is present
or not.

The already limited number of PNe available for a given galaxy often
results in a problem of small statistics. At the same time, the
binning procedure reduces the data to only a few points, of which
only those corresponding to the  brightest PNe are used for the
fit, and thus the issue of small number statistics is only aggravated.

A cumulative PNLF has the advantage of improving the statistics by
using more points to do the fit, while the assumptions about the
histogram bins (e.g. size and number) are no longer necessary.

The cumulative PNLF can be obtained as
\begin{equation}
\label{eq:antidev}
I(m;N_T,m^*)= \int^{m}_{m^*}N(m';N_T,m^*)dm',
\end{equation}
Using equations~(\ref{eq:pnlf2}) and~(\ref{eq:antidev}) one obtains, 
\begin{eqnarray}
I(m;N_T,m^*)&=&N_T[A\,e^{3m^*-B m}+C\,e^{0.307m} \nonumber \\
            & &-(A+C)\,e^{0.307m^*}] ,
\label{eq:cumulative}
\end{eqnarray}

where, $A=0.37133$, $B=2.693$ and $C=3.25733$.
An example of this cumulative PNLF can be found in Pe\~na et al. (2007), 
where it was obtained for a sample of 20 PNe from NGC\,3109.\\
If the cumulative PNLF is used, the uncertainties introduced by the
binning procedure are eliminated, however there is still the issue of
restricting the fit to the brightest PNe in the sample.
Certainly at the faintest end the PNe sample is incomplete, and 
the luminosity function will have a drop in number (a plateau if the
cumulative function is considered). 

In star-forming galaxies where a dip is present, it is not easy to fit the PNLF (standard or  cumulative), and what 
is usually done is to restrict the fit to the brightest PNe, before the dip 
in the luminosity function. Results obtained for these  galaxies agree (within 
the error bars) with the canonical value of  $M^*_{5007}$ . If the dip is due to a second stellar population, the underlying assumption is 
that this second one does not overlap with the first population of 
the bright end of the PNLF, at least in the portion used to fit the data.

\section{The two-mode planetary nebula luminosity function}
Since the number of observable PNe  in extragalactic sources is rather
limited, we would like to use as many objects as possible to
characterise their luminosity function. In this regard the cumulative
PNLF seems the most natural choice. Instead of restricting the analysis to 
the brightest PNe in the sample, we propose to explicitly include a second mode in
the fitted luminosity function. Of course, any new two-mode luminosity function 
must have the same properties as the canonical PNLF in order to reproduce the 
results for galaxies with a single mode. One should point out
that using a cumulative function, the data fit may have some inconsistencies due to lack of
statistical independence of the cumulative data.\\

\noindent
Thus, the proposed two-mode PNLF is:
\begin{equation}
\label{eq:2pnlf}
N(m)=N(m;N_{T1},m^*_1)\times \mathrm{H}(m-m_{cut})+N(m;N_{T2},m^*_2)   
\end{equation}
where, $\mathrm{H}(m-m_{cut})$ is the ``Heaviside step function'', defined as
\begin{equation}
\label{eq:heaviside}
\mathrm{H}(m-m_{cut})=\left\{
\begin{array}{lc}
1,  &  m\leq m_{cut}, \\
0,  &  m > m_{cut}.
\end{array}
\right.
\end{equation}
The proposed function is the sum of two standard  PNLFs, allowing each
mode to have a different $N_T$, and $m^*$. One of the two modes is
truncated abruptly at a magnitude ($m_{cut}$), this is a simple mathematical artifact, certainly, one could introduce another cutoff for the second population, or extend the function to
three or more modes. However, in the spirit of having as few free
parameters as possible we will adopt the form in equation
(\ref{eq:2pnlf}). 

A cumulative function of the two-mode luminosity function of equation
(\ref{eq:2pnlf}) can be obtained by integrating over the magnitude. Integration of the first mode gives
\begin{eqnarray}
I_1(m;N_{T1},m^*_1,m_{cut}) & = &\int^{m_{up}}_{m_1^*}N(m';N_{T1},m^*_1)dm',
\label{eq:I1a} \\
 & = & N_{T1}[A\,e^{3m^*_1-B m_{up}}+C\,e^{0.307 m_{up}}\nonumber   \\
                &  &-(A+C)\,e^{0.307m^*_1}], 
\end{eqnarray}
where the integration is stopped at $m_{cut}$ by virtue of
\begin{equation}
m_{up}=\min(m,m_{cut}) \geq m_1^*.
\end{equation}

Notice that after the cutoff magnitude $m_{cut}$ the first luminosity
function drops to zero, but the cumulative remains at a constant value.
The integration of the second mode is analogous to the sample  population
case (for $m \geq m^*_2$),
\begin{eqnarray}
\label{eq:I2}
I_2(m;N_{T2},m^*_2)&= & N_{T2}[A\,e^{3m^*_2-B m}+C\,e^{0.307 m}\nonumber   \\
                &  &-(A+C)\,e^{0.307m^*_2}].
\end{eqnarray}
The total cumulative luminosity function is then the sum of
the two modes:
\begin{eqnarray}
\label{eq:2cumulative}
&N_c(m;N_{T_1},m_1^*,m_{cut},N_{T2},m_2^*)=\nonumber \\
& I_1(m;N_{T2},m^*_1,m_{cut})+I_2(m;N_{T2},m^*_2).
\end{eqnarray}

\section{Results}

We have used observations of nine galaxies described below and  constructed  their
cumulative  PNLFs (Table~1). 
 We then use the  {\sc aga-v1} code
(Rodr\'iguez-Gonz\'alez et al. 2012) for each galaxy to obtain the
best fit of one- and two-mode cumulative  PNLFs. 
The {\sc aga-v1} code uses the Asexual Genetic Algorithm  described in
Cant\'o et al. (2009), and allows to find the
best fit exploring a wide parameter range  in order to minimize a
  merit function, in this case a $\chi^2$. The code varies
simultaneously and independently all the parameters of the fit in such a
space.
 
To estimate the uncertainty in our fitting procedure, we perform 100
realizations for every PNLF in each galaxy presented in this paper.
Each realization is obtained varying the original data as follows: 
\begin{equation}
\label{yiprim}
y'_i=y_i+\sigma_i \xi_i, 
\end{equation}
where, $y'_i$ represents the i-th data point of the new set, $y_i$
  and $\sigma_i$ are the original data and its associated  
errors, which are assumed to follow a Poisson distribution,
\begin{equation}
\label{err}
\sigma_i=\sqrt{y_i},
\end{equation}
and $\xi_i$ is a uniformly distributed random deviate.
Each data set constructed is fed to the {\sc aga-v1} code and yields a
set of fitting parameters that minimize the $\chi^2$ merit function.
We allow the code to find $m_1^*$, $m_2^*$,
and $m_{cut}$ anywhere from half a magnitude below the minimum in
the sample to half a magnitude above the maximum in the sample,  
the range covered for $N_{T,1}$ and  $N_{T,2}$ is from $10^{-7}$ to
$10^2$. An additional constraint that we have enforced is  
that $m_{cut}\geq m_1^*$, otherwise the results are unphysical. It is
important, however, to mention that $m_1^*$ and $m_2^*$  are allowed
to be smaller or larger to each other.

The {\sc aga-v1} code is similar to many Monte-Carlo methods, where
an ensamble average is used to obtain the average values and
dispersion of the parameters of the fit.
Thus, from the 100 realizations for each galaxy we obtain the fit
parameters from the average value of each of the parameters ($N_{T1}$,
$m*_1$, $m_{cut}$, $m*_2$, and $N_{T2}$);  and an estimate of the uncertainty
(from standard deviation). 

In addition we compute a Kolmogorov-Smirnov (K-S), to asses
the likelihood of the data and proposed function
(eq.~\ref{eq:2cumulative}) to arise from the same distribution.

The K-S test gives two measures of the goodness of the fit: a
significance level with a value between 0 and 1, listed simply as
K-S in Table 3 (a value close to 0 means that the data and the
function are significantly different), and the maximum departure
(listed as D) between the cumulative function of the  data and the
function provided.

 C89, Herrmann et al. 2008, Feldmeier et al. 1997, also used a K-S
test to compare their fits (from a maximum likelihood method) to observations.

Our best fits are presented in Table~2, where 
we named the galaxies, and describe for each of them the best fits parameters ($N_{T,1}$, $m_1^*$, $m_{cut}$ 
$N_{T,2}$ and $m_2^*$. ) for the cumulative PNLF. In Table~3 we show the results for these tests in our best fits (presented in Table~2), the number of the galaxy, 
the total number of PNe and the number of PNe used for each fit, the K-S and D values for the test and the number of modes in the cumulative PNLF are presented.

\subsection{The sample of galaxies}

In order to explore the PNLF for galaxies of differents Hubble-types,
we selected 9 galaxies: four irregulars (SMC, LMC, 
NGC\,6822, NGC3109), three spirals (NGC\,300, M31 and M33), a dwarf elliptical 
(NGC\,205), and an elliptical galaxy (NGC4697).\\

In Table~1,  we describe the sample by: name, Hubble-type of the galaxy, number of PNe observed, 
and two fit parameters of the PNLF ($N exp(-0.307\mu) $ and $m^*_{5007}$). For some galaxies, we 
show two fit parameters values, because the authors that present de PNe sample do not fit a PNLF. 
For NGC\,205, we take the inside sample of PNe, since Corradi et al. (2005) discuss that the other ones are very faint.   

\begin{table}
\centering
\caption{Sample of galaxies with Hubble-type, number of PNe and some previous fit values.
 $^a$~Reid \& Parker (2010), $^b$~Jacoby \& De Marco (2002),  $^c$~Hern\'andez-Mart\'inez \& Pe\~na (2009), $^d$~Pe\~na et al. (2007), $^e$~Pe\~na et al. (2012), $^f$~Ciardullo et al. (2004), $^g$~Ciardullo et al. (2002), $^h$~Corradi et al. (2005), $^i$~M\'endez et al. (2001).}
\begin{tabular}{lcccccc}
\hline
Galaxy          &Type  & N$\mathrm{_{PNe}}$  &$N exp(-0.307\mu) $                 & $m^*_{5007}$  \\
                     &          &                                    &                                                  &    (mag)       \\            
\hline
LMC              & Ir       &    164$^{(a)}$            &   -                                               & 14.05$^{(a)}$  \\
                      &          &                                  & 2.87$\times10^{-1}$$^{(d)}$      & 14.23$^{(d)}$\\
SMC             & Ir       &     59$^{(b)}$              &  -                                                 & 14.8$^{(b)}$   \\  
                      &          &                                  & 9.31$\times10^{-2}$$^{(d)}$      & 14.82$^{(d)}$\\ 
NGC\,6822   & dIr     &     23$^{(c)}$              &(5.0$\pm$0.6)$\times10^{-3}$$^{(c)}$  & 20.43$\pm$0.19$^{(c)}$  \\   
NGC\,3109   & dIr    &     20$^{(d)}$               &2.62$\times10^{-3}$$^{(d)}$                 & 21.18$^{(d)}$ \\
NGC\,300     & Sp    &    100$^{(e)}$            &(8.9$\pm$ 0.52)$\times10^{-3}$$^{(e)}$& 22.44${\pm}$0.19$^{(e)}$\\
M\,33            & Sc    &    152$^{(f)}$             &                -                             & 20.39$^{+0.07}_{-0.11}$$^{(f)}$  \\
M\,31            & Sb    &    298$^{(g)}$           &1.5$\times10^{-1}$$^{(d)}$     & 20.20$^{(d)}$    \\
NGC\,205     & Sph  &     35$^{(h)}$            &             -                             & 20.17$\pm$0.21$^{(h)}$ \\
NGC4697     & E6    &   535$^{(i)}$             &           -                               & 25.63$\pm$0.18$^{(i)}$   \\
\hline \hline
\end{tabular}
\end{table}

\begin{figure}
\centering
\includegraphics[width=\columnwidth]{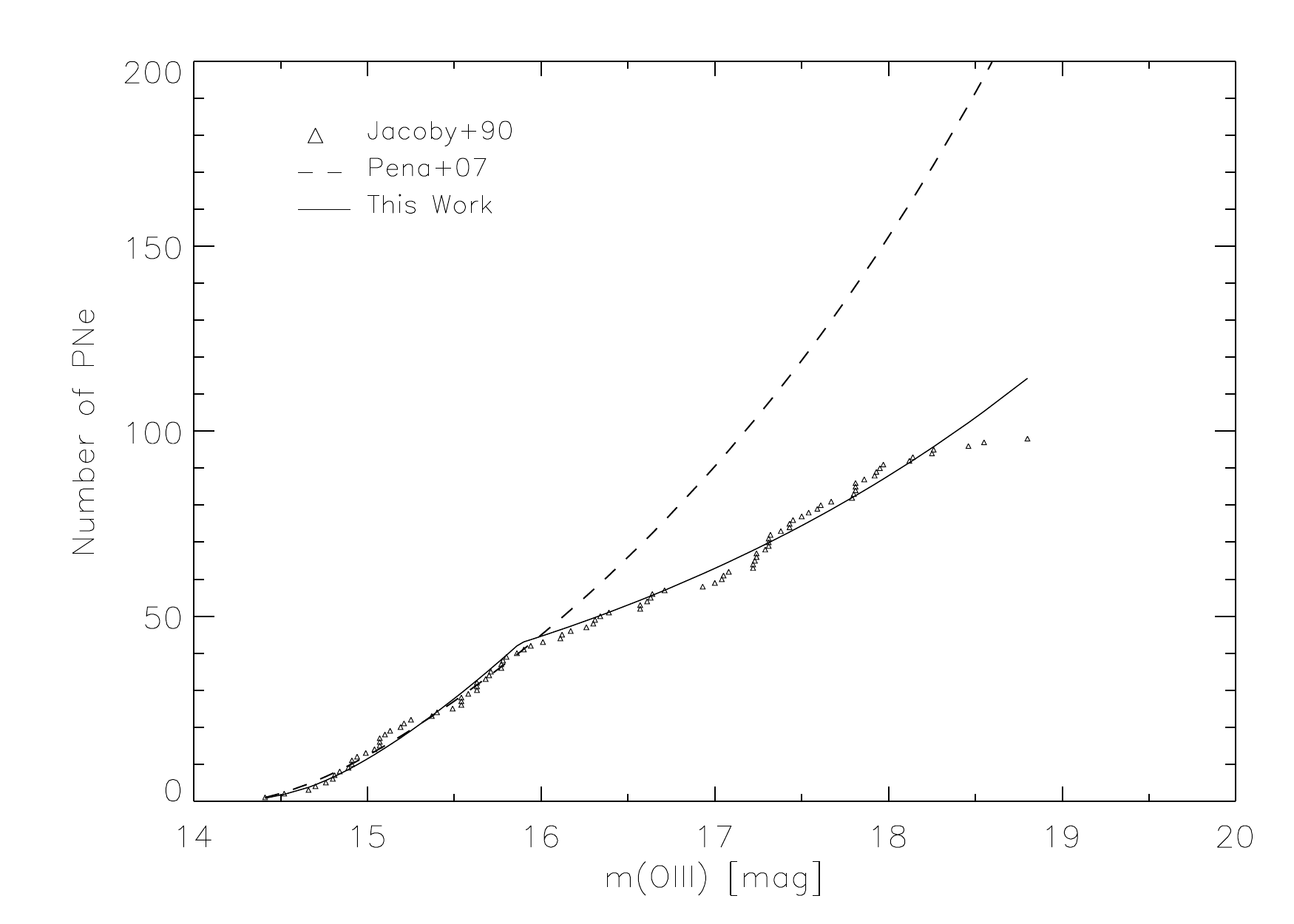}
\caption{Cumulative PNLF of the Large Magellanic Cloud. The
  \emph{triangles} are the data in Reid \& Parker (2010) and the
  \emph{solid} line is our two mode fit.} 
\label{f:LMC}
\end{figure}

For LMC, we take the sample of Reid \& Parker (2010), in Figure~\ref{f:LMC} we show
 the observations and the fit to the cumulative PNLF
in the LMC. The ``triangles'' are the observational data presented by Reid \& Parker (2010) and the solid line is our best fit using the two modes cumulative PNLF (equation~\ref{eq:2cumulative}, hereafter 2mc-PNLF).\\ 

Our best fit yields $m^*_{1}=14.18\pm0.14$, 
$N_{T1}$=(1.71$\pm$0.04)$\times$10$^{-1}$, $m_{cut}$=15.95$\pm$0.4, 
$N_{T2}$=(1.19$\pm$0.01)$\times 10^{-1}$ and 
$m^*_2$=15.62$\pm$0.76.  A Kolmogorov-Smirnov test shows that the
2mc-PNLF is not inconsistent with
the data (the value of the K-S paramters are listed also in Table~\ref{tab:ks}).
The value of $m^*_1$ is in very good agreement with that reported
by Pe\~na et al. (2007) for  $m^*_{5007}$, while the $N_{T1}$
obtained by the 2mc-PNLF is lower than the $N_T$ for the simple
cumulative PNLF.
Naturally the reason for this is that the total population of PNe
is divided into two populations  as  proposed in
equation~\ref{eq:2cumulative}.
$N_T$ is related with the number of planetary nebula expected at a
certain magnitude $m$. $N_{T1}$ and $N_{T2}$ are related with the
number of PNe in the first and in the  second mode (or population),
respectively. 
From the   values of $m^*_1$, $m^*_2$ and $m_{cut}$, we know that the PNe for the
first mode of the function (the brightest one) are overlaping with the second
mode in the $\sim 14.56$ to $\sim 15.87$~mag range. Therefore, $N_T$
includes a contribution from the second population. Thus, one should not expect $N_T$ and $N_{T1}$ to
coincide in general, but rather  $N_{T1} \le N_T$. \\

\begin{figure}
\centering
\includegraphics[width=\columnwidth]{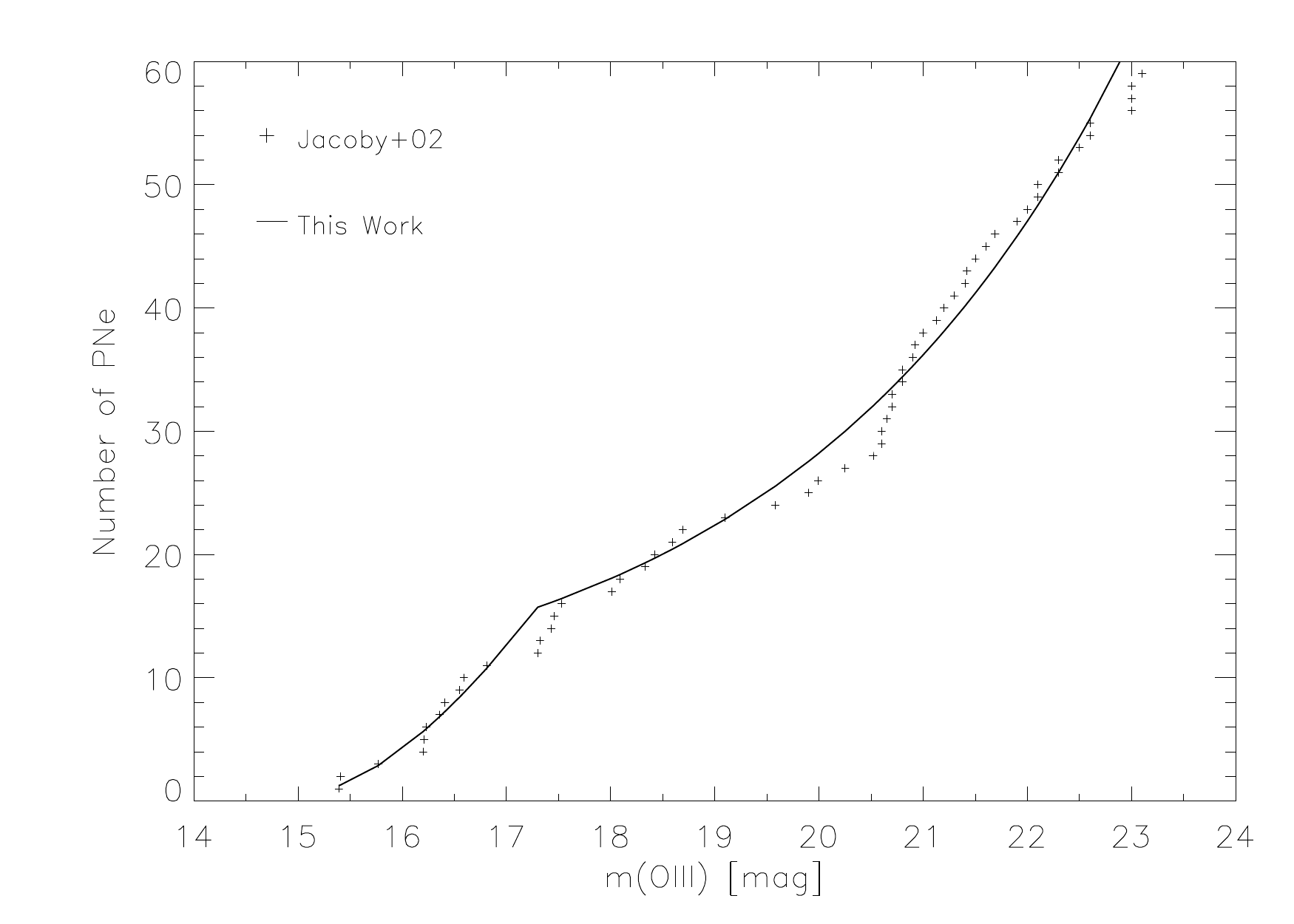}
\caption{Cumulative PNLF of the Small Magellanic Cloud. The
  \emph{plus symbols} are the data in Jacoby \& De Marco (2002), 
the \emph{solid} is our two mode fit.} 
\label{f:SMCJac}
\end{figure}

 For the SMC,  Jacoby \& De Marco (2002) obtained a statistically complete PNLF
in 10~mag (from 14 to 24). Their figure~6 shows a dip, starting
at a magnitude of 17, and ending at 20~mag. The cumulative PNLF is shown in Figure~\ref{f:SMCJac} with the crosses. The best fit using 
our 2mc-PNLF (continuous line) gives $N_{T1}=$3.97$\times 10^{-2}$, $m^*_1=14.82$ and 
$m_{cut}=17.29$ for the first mode and $N_{T2}=1.46$$\times 10^{-2}$ 
and $m^*_2=15.65$ for the second one. Our $m_{cut}$(=17.29) fits very
well with the dip seen in the observed PNLF.

\begin{figure}
\centering
\includegraphics[width=\columnwidth]{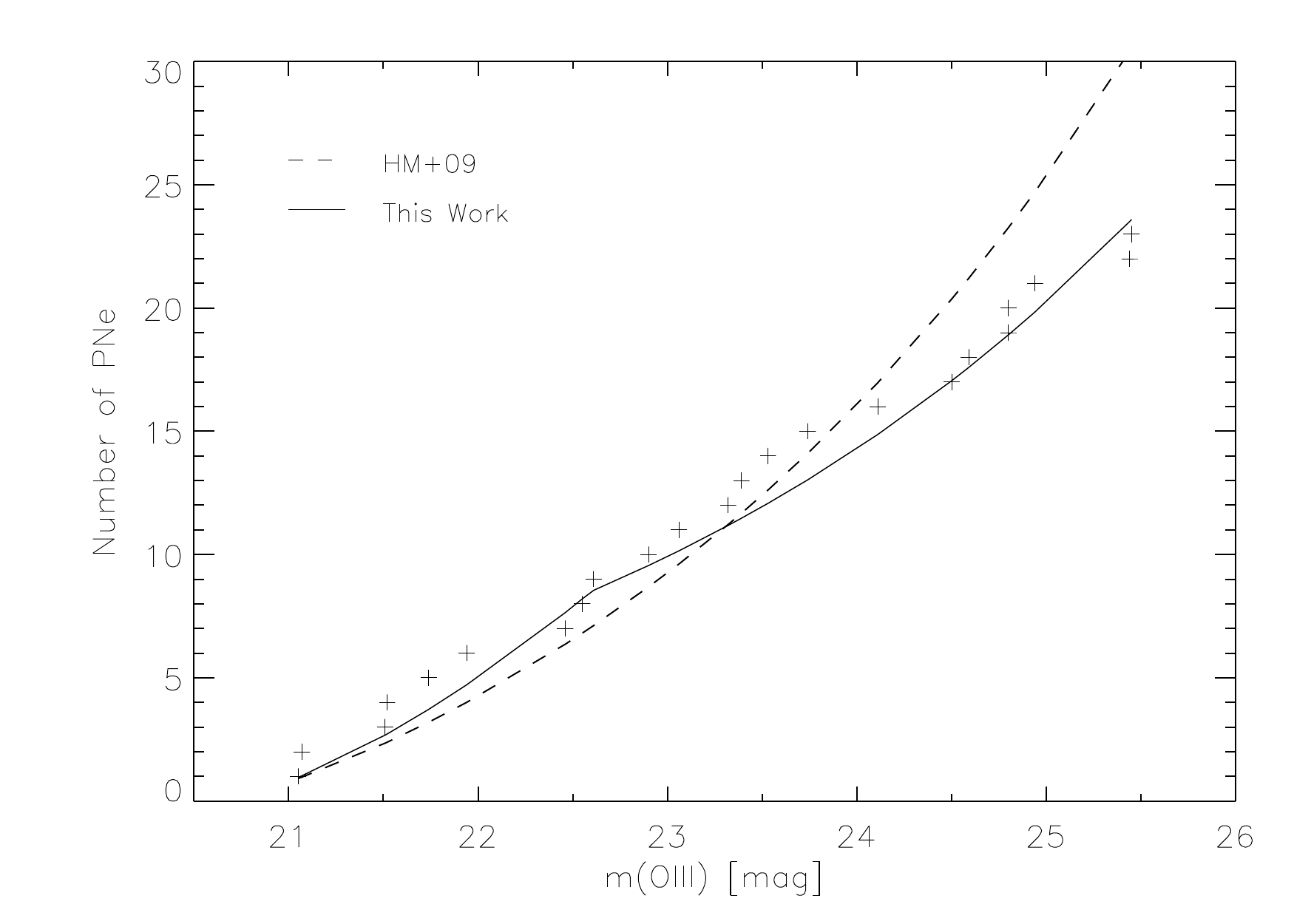}
\caption{Cumulative PNLF of NGC\,6822. The \emph{plus} signs and the
  \emph{dashed} line, are the data and the fit of HMP09, respectively. The
  \emph{solid} line is our two mode fit.}
\label{f:6822}
\end{figure}

\begin{figure}
\centering
\includegraphics[width=\columnwidth]{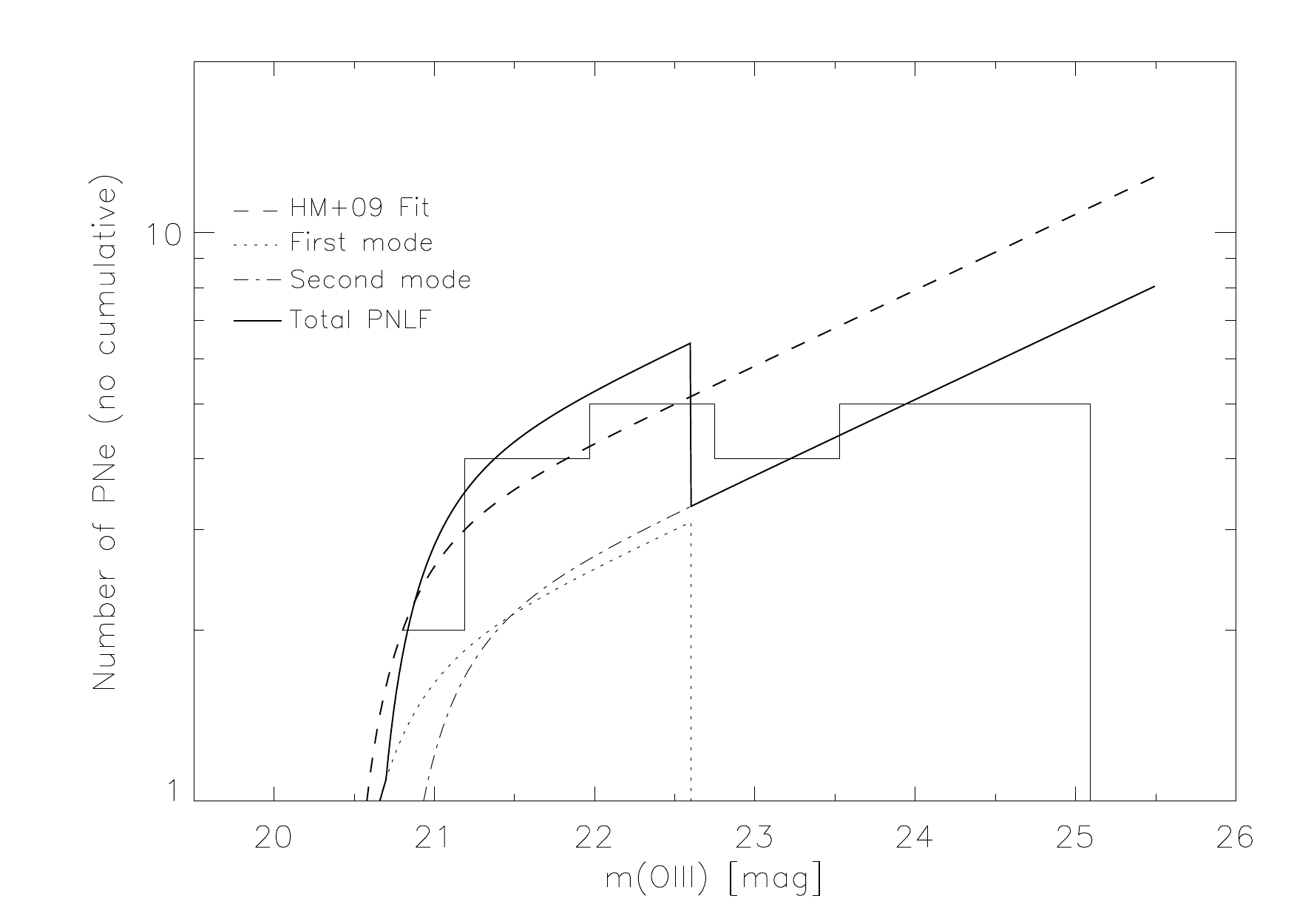}
\caption{PNLF (non-cumulative) of NGC\,6822. The \emph{thin solid} line and the
  \emph{dashed} line, are the histogram of the observational data and the fit of HMP09, respectively. 
The \emph{dotted} and \emph{dashed-dotted} lines are the first and the second mode
of our fitted function and  the
  \emph{thick solid} line is our two mode fit.}
\label{f:6822Can}
\end{figure}

Figure~\ref{f:6822} shows the cumulative luminosity function of the 
PNe in NGC\,6822. The  \emph{plus} symbols and the \emph{dashed} lines are the
observational data and the empirical PNLF presented by HMP09. Their best fit 
is given by $N_T=5\times 10^{-3}$, $m^*=20.43$, and a break in the
PNLF distribution at a magnitude of  $\sim$22.65. 
They related the break (dip in the non cumulative PNLF) with a second
population in this galaxy. The best fit using 
the 2mc-PNLF yields $N_{T1}=3.01\times 10^{-3}$, $m^*_1=20.37$,
$m_{cut}=20.6$, $N_{T2}=3.21\times 10^{-3}$ and
$m^*_2=20.7$.
In Figure~\ref{f:6822Can} we show the non-cumulative luminosity  
functions for the same galaxy, the \emph{dashed} line is the empirical PNLF and  
the solid line is the non-cumulative PNLF obtained with the parameters of
the PNLF fit. The \emph{dotted} and \emph{dash-dotted} lines are
the individual modes of the PNLF. The PNLF fits  well the
observational data of NGC\,6822. Note that the locus and the depth of
the dip are reproduced with our fit.

\begin{figure}
\centering
\includegraphics[width=\columnwidth]{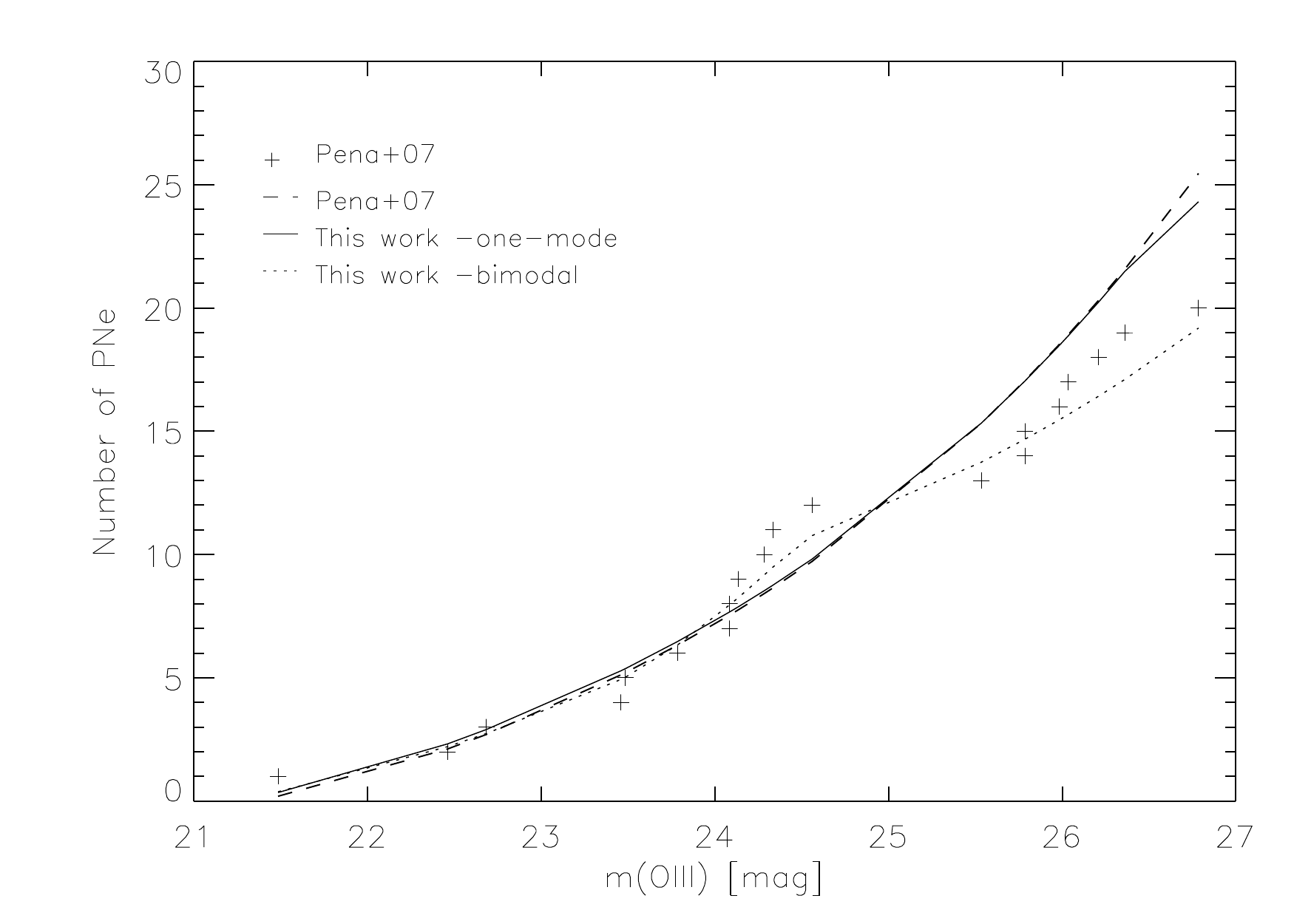}
\caption{Cumulative PNLF of NGC\,3109. The \emph{plus} signs and the \emph{dashed} line, are the observational data  and  the single mode fitted in Pe\~na et al. (2007). The \emph{solid} line
  and the \emph{dotted} lines, are the one and two mode best fits,
  respectively,  obtained with our genetic algorithm.}
\label{f:3109}
\end{figure}

NGC\,3109 has a peculiar PNLF, which is shown in
Figure~\ref{f:3109}.  What is particular about this galaxy is the void in
its luminosity function around $m_{O III}\sim 25$, which could be
identified as the signature of a second PN population.
However, the fitting algorithm applied to the PNLF of NGC\,3109 yielded a
solution consistent with a single mode (the algorithm yielded a second 
mode with $m_2^*\sim m_{cut}$, the fit parameters are given in Table~\ref{tab:fit}).
This `effective' single mode was somewhat worrisome since there is a
noticeable plateau in the cumulative function at $m_{O III}\sim 25$.
For this reason we re-ran the fitting algorithm, firstly  with a single population, 
and secondly with two modes, but arbitrarily restricting $m_{cut}$ to lie
in the range of 23.5-25.5 mag.  

In order to explore possible correlations  between the fitting parameters, we calculate the covariance
matrix (see Feigelson \& Babu 2012),
\begin{equation}
\label{eq:covariance}
q_{ij}=\frac{1}{N-1}\sum^N_{k=1}=(c_{i,k}- <c_i>)(c_{j,k}-<c_j>)
\end{equation}
where, $c_i$ and $c_j$ are the fitting parameters (i.e. $N_{T1}$, $m*_1$, $m_{cut}$ , $N_{T2}$, $m^*_{2}$) and 
the sum is carried out over 100 realizations ($N=100$). $<c_i>$ and $<c_j>$ are the mean values obtained over
the realizations (see Table~\ref{tab:fit}). Using the mean values  of the fitting parameters presented in row 5 of
Table~\ref{tab:fit} and 
the fitting parameters obtained for each of the realizations we calculated the covariance matrix PNLF of NGC\,3109,

\[q_{ij}=  \left( \begin{array}{ccccc}
         0.000    &     0.000  &      0.000    &    0.000  &       0.000 \\
         0.000    &    0.250   &     -0.090     &    0.000   &      0.024 \\
         0.000    &    -0.090  &       0.129   &      0.000  &      -0.047 \\
         0.000    &   0.000   &      0.000    &     0.000   &       0.000 \\
         0.000    &     0.024   &     -0.047  &      0.000   &      0.323 \\
 \end{array} \right)\] 

One can see that the covariance matrix has cross-correlations
close to zero (given by the off-diagonal matrix coefficients). These small values indicate that pairs
of fitting parameters are not linearly correlated. Moreover, orthogonality tests for the fitting parameters
show that $N_{T1}$ and $N_{T2}$ are orthogonal and  all pairs of fitting parameters have angles close to $90^\circ$.

\begin{table*}
\centering
\caption{Cumulative Planetary Nebula Luminosity Function fits}
\label{tab:fit}
\begin{tabular}{lcccccc}
\hline
Galaxy & $N_{T1}$   & $m^*_{1}$ & $m_{cut}$  & $N_{T2}$ & $m^*_{2}$ \\
\hline
LMC & (1.71$\pm$0.04)$\times 10^{-1}$& 14.18$\pm$0.14  &  15.95$\pm$0.40 &(1.19$\pm$0.01)$\times 10^{-1}$   & 15.62$\pm$0.76 \\
SMC       & (4.86$\pm$0.01)$\times 10^{-2}$& 14.65$\pm$0.05  &  16.81$\pm$0.12 &(3.39$\pm$0.01)$\times 10^{-2}$& 15.24$\pm$0.64 \\  
NGC\,6822 & (3.01$\pm$1.02)$\times10^{-3}$ & 20.37$\pm$0.12 & 22.60$\pm$0.54 & (3.21$\pm$0.02)$\times 10^{-3}$ & 20.70$\pm$0.15\\
NGC\,3109 & (2.58$\pm$0.52)$\times 10^{-3}$ & 21.05$\pm$0.35 &
26.67$\pm$1.20 & (1.68$\pm$0.07)$\times 10^{-1}$& 28.04$\pm$1.30 \\
NGC\,3109$\dag$   & (2.35$\pm$0.03)$\times 10^{-3}$ & 20.99$\pm$0.06 & --- & --- & --- \\
NGC\,300   & (9.23$\pm$0.05)$\times 10^{-3}$ & 22.66$\pm$0.03 & 27.62$\pm$0.26&(2.69$\pm$1.71)$\times 10^{-1}$ & 28.48$\pm$0.56\\
M\,31& (6.49$\pm$0.04)$\times 10^{-2}$ & 20.24$\pm$0.01 & 22.29$\pm$0.6 & (8.66$\pm$0.08)$\times 10^{-2}$ & 22.17$\pm$0.49\\
M\,33    & (5.6$\pm$0.02)$\times 10^{-2}$ & 20.46$\pm$0.06 & 22.89$\pm$0.12&(2.83$\pm$0.37)$\times 10^{-2}$ & 22.95$\pm$0.30\\
NGC\,205     & (8.19$\pm$0.02)$\times 10^{-3}$ & 20.19$\pm$0.04 & 24.29$\pm$0.10&(3.37$\pm$1.37)$\times 10^{0}$ & 27.80$\pm$1.33\\
NGC4697    &  (6.52$\pm$0.18)$\times 10^{-2}$ & 25.51$\pm$0.01&   ---             &     ---     &   ---\\
\hline \hline
\end{tabular}
\vskip 1cm
\end{table*}

\begin{figure}
\centering
\includegraphics[width=\columnwidth]{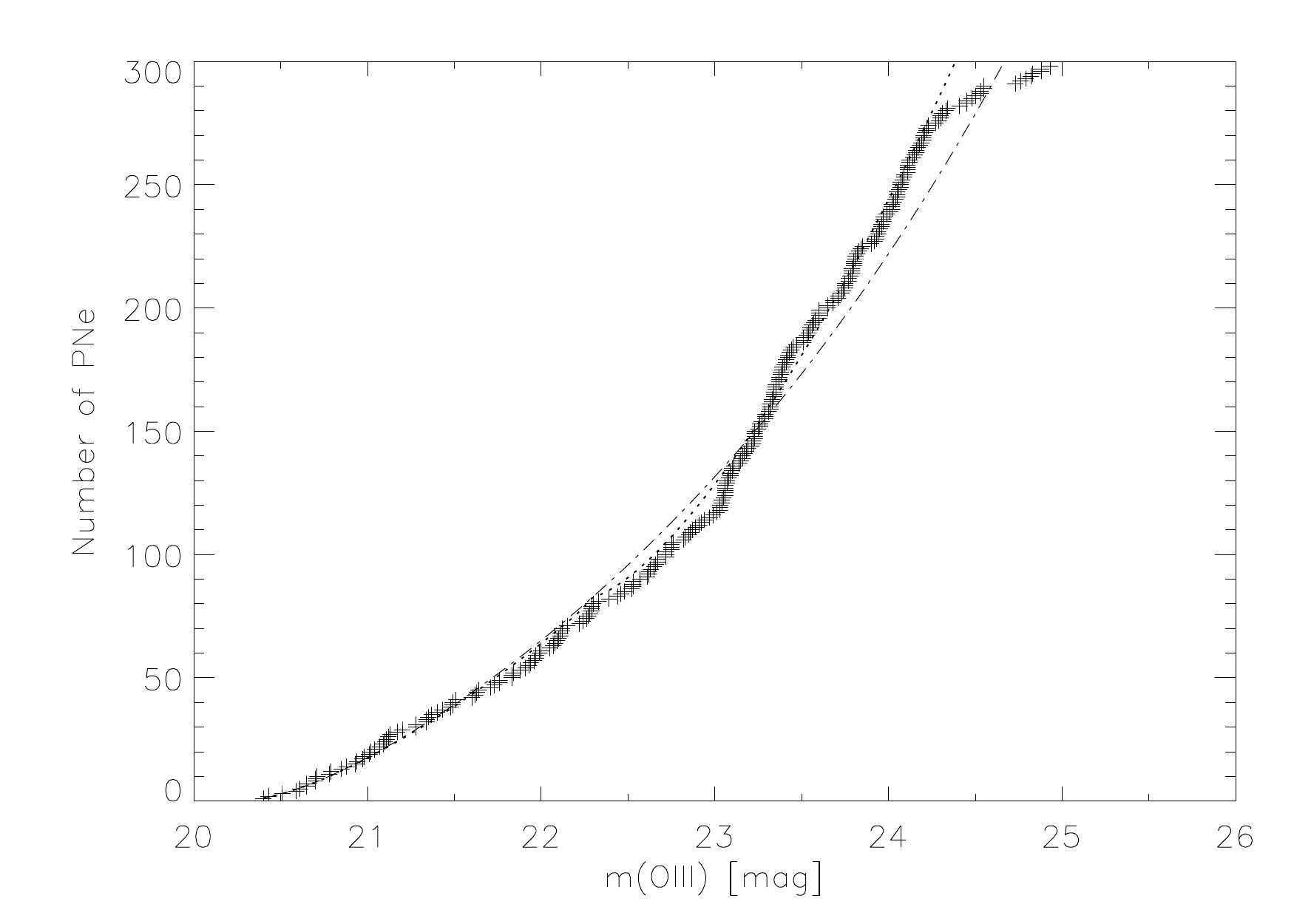}
\caption{Cumulative PNLF of M\,31.}
\label{f:m31}
\end{figure}

\begin{figure}
\centering
\includegraphics[width=\columnwidth]{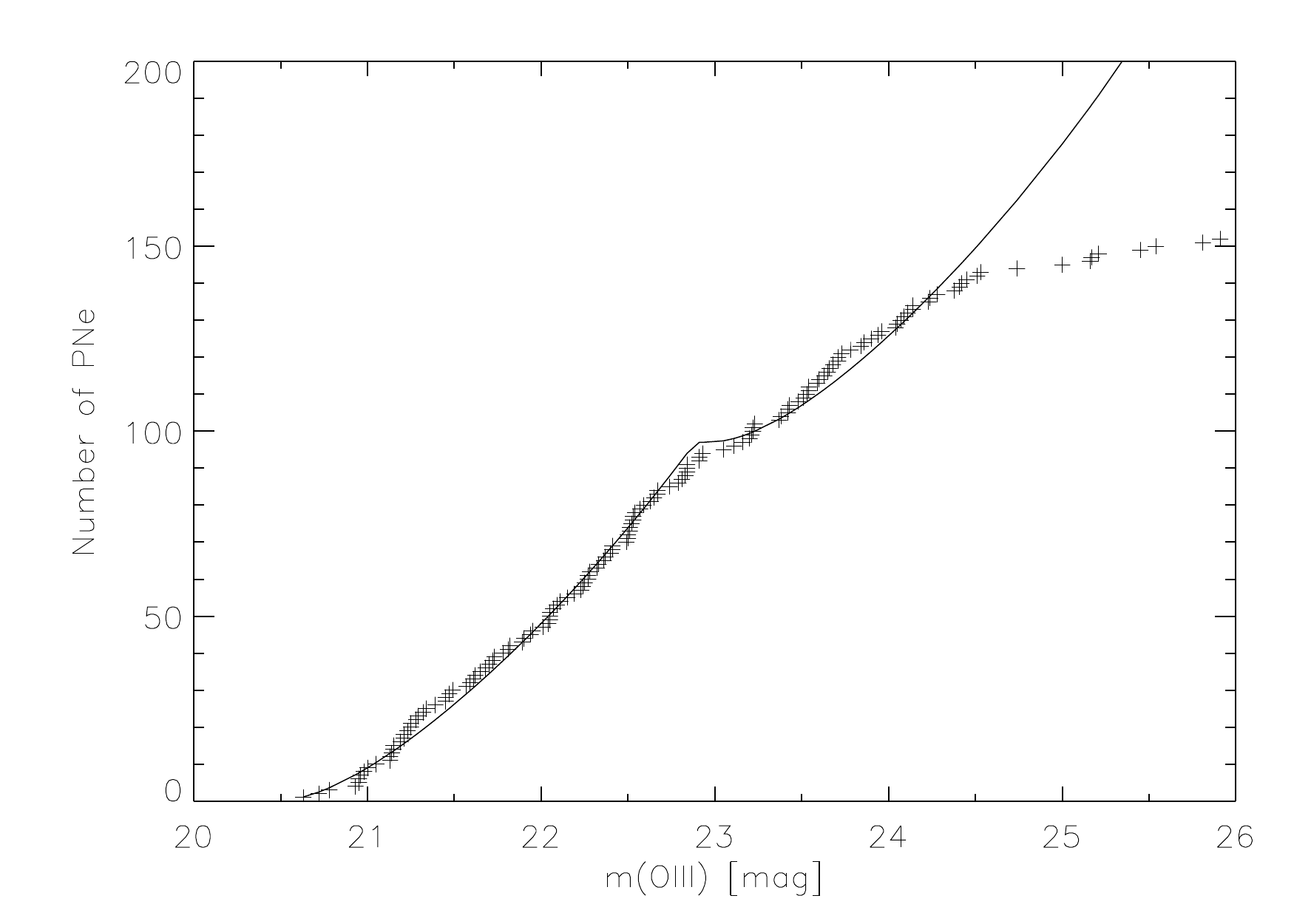}
\caption{Cumulative PNLF of M\,33.}
\label{f:m33}
\end{figure}

\begin{figure}
\centering
\includegraphics[width=\columnwidth]{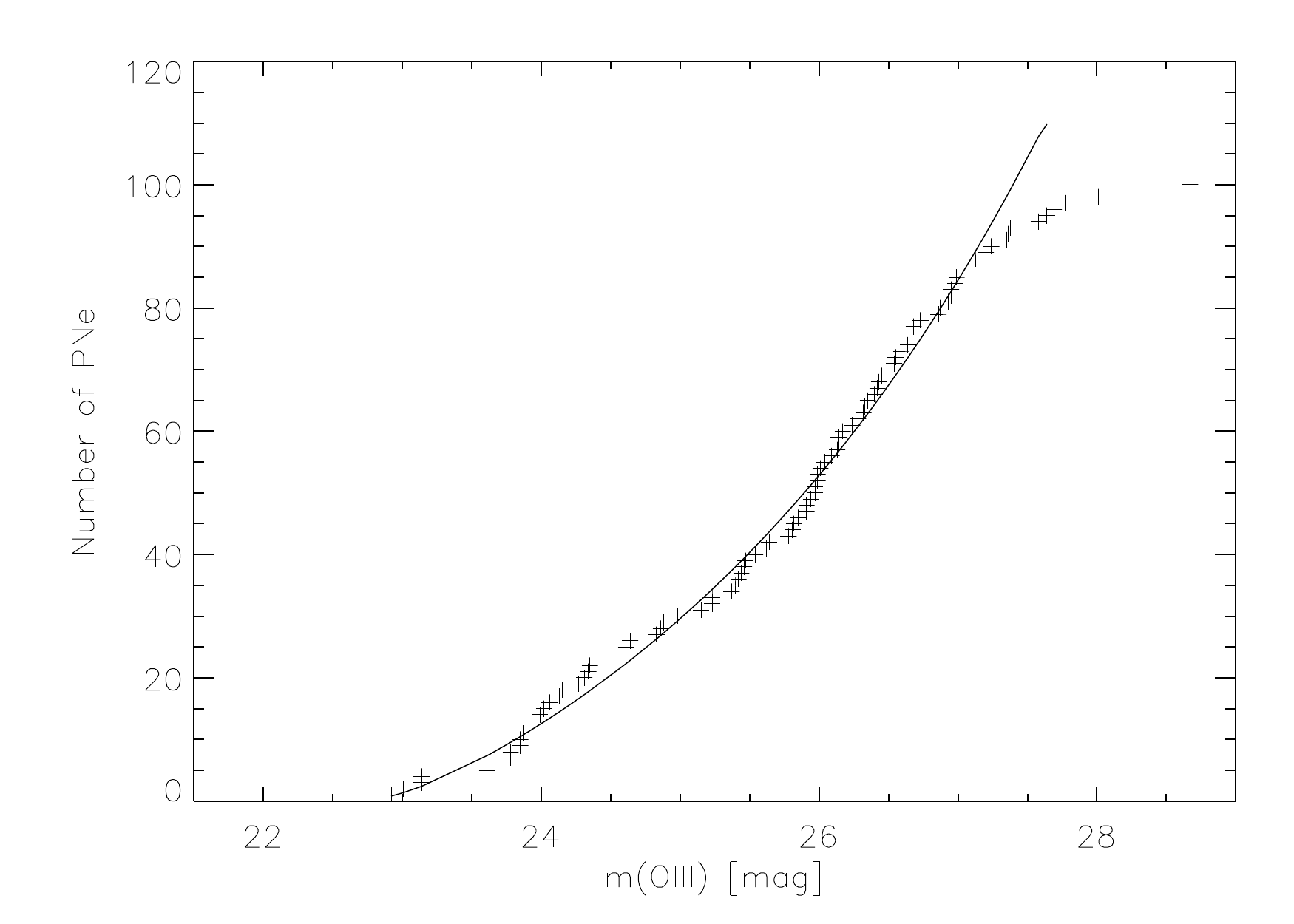}
\caption{Cumulative PNLF of NGC\,300.}
\label{f:300}
\end{figure}

Figures \ref{f:m31}, \ref{f:m33} and \ref{f:300} show the cumulative 
PNLF for the spiral galaxies M\,31, M\,33 and NGC\,300,
respectively. For M\,31 and M\,33 the PNLF is consistent with two
modes, but the shape of the luminosity function for these two galaxies
is somewhat different. For M\,31, the PNLF has a 
noticeable steepening of the slope after a magnitude $\sim$22.5. The 
cutoff magnitude of the first population obtained for this galaxy is 22.29,
 while the second mode starts of at 22.17. Thus, in these galaxies the overlap
of the two populations is rather marginal.

For M\,33, the PNLF (Figure~\ref{f:m33}) has a distinct break at a
magnitude $\sim 22.8$. This is possible  if the two populations do not overlap, as
found from the fit, in which $m^*_2> m_{cut}$.  For instance in figure
6 of Ciardullo et al. (2004) one could see evidence of two dips in
their PNLF. In the present work we only confirm one dip at 22.9~mag,  
in accordance with their second dip, which is the sharper.

Figure~\ref{f:300} shows the PNLF 
of NGC\,300 (a spiral galaxy). In our best fit the second 
mode of the PNLF starts very close to the completeness limit of the 
observational data, meaning that it is consistent with a single mode PNLF.
However, the $N_T$ fit by Pe\~na et al (2012) is different from the value obtained
using  the present form of the  PNLF. Again, as in the case of NGC\,3109, NGC\,300 has to be 
analyzed and improve de sample, basically by completing the whole galaxy, 
in order to explore the shape of the PNLF in more detail. 

\begin{figure}
\centering
\includegraphics[width=\columnwidth]{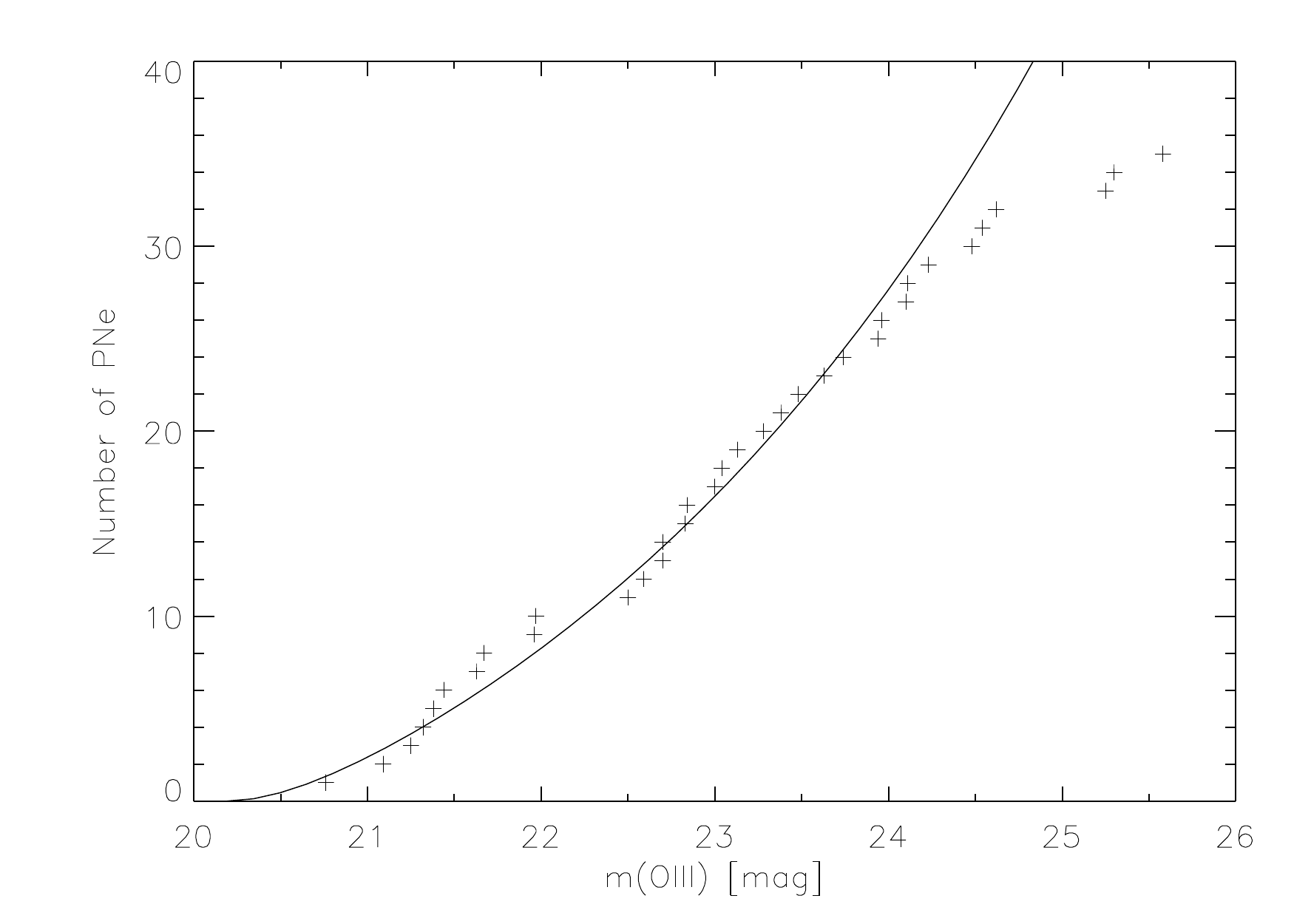}
\caption{Cumulative PNLF of NGC\,205.}
\label{f:205}
\end{figure}

In Figure~\ref{f:205} show the PNLF for the dwarf elliptical galaxy. 
NGC\,205 has a similar behavior  as  NGC\,300, in which a second
population was found to begin at a magnitude well within the region in which the
sample is incomplete. Therefore, it can be considered  as consistent with a single
mode PNLF.

\begin{figure}
\centering
\includegraphics[width=\columnwidth]{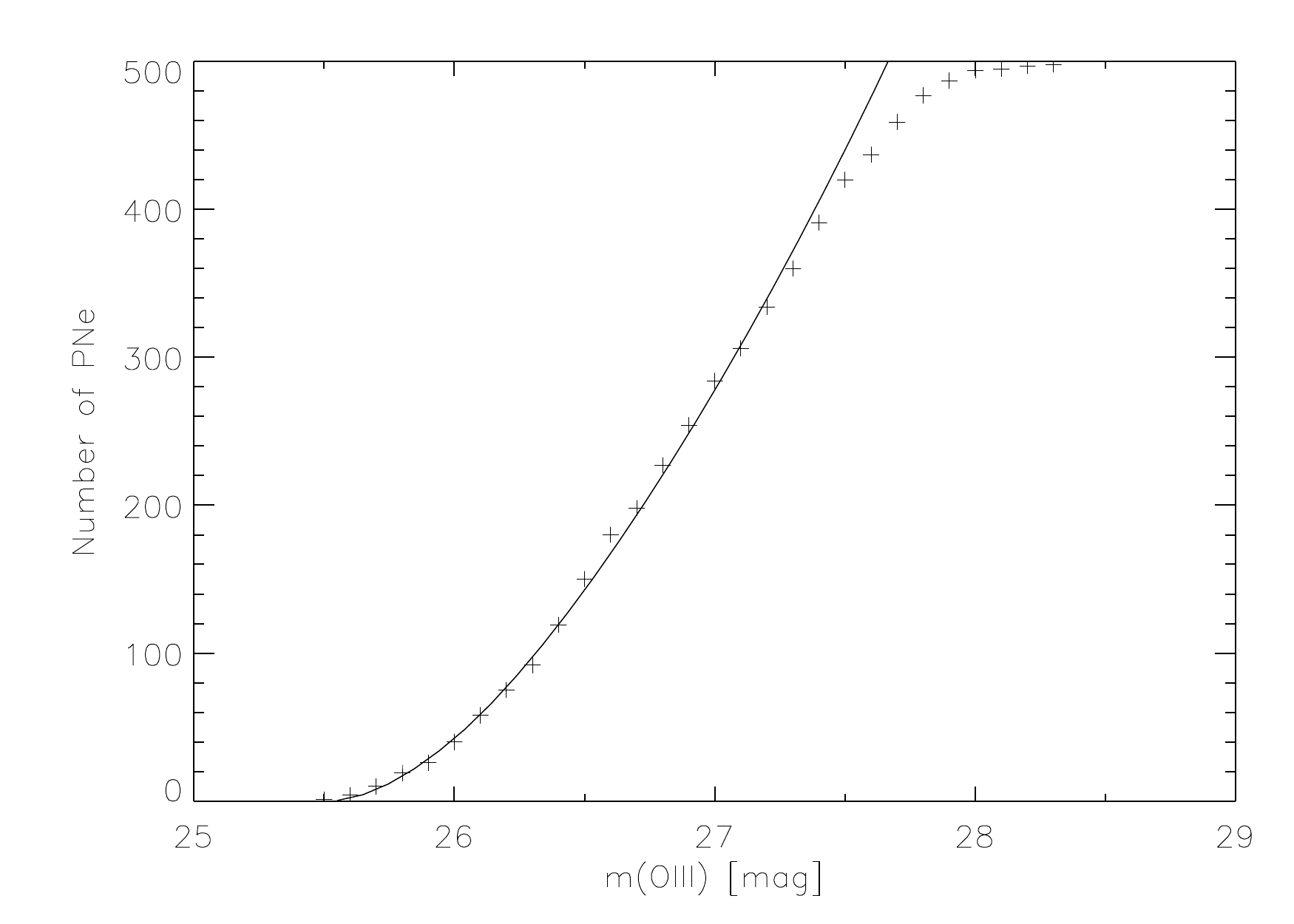}
\caption{Cumulative PNLF of NGC\,4697.}
\label{f:4697}
\end{figure}

Finally, Figure~\ref{f:4697} shows the PNLF for the elliptical galaxy NGC\,4697. This data were obtained from 
M\'endez et al. (2001), they argued the possibility of a dip, but with our method we do not support that idea. We clearly see only a one-mode PNLF. 

As far as  we can say from  our study, It is not clear if the number of modes correlates directly
with the Hubble type of the galaxy. But we can see that in most of the galaxies with recent star formation  
that it is present a dip.

\begin{table}
\centering
\caption{Kolmogorov-Smirnov test results}
\label{tab:ks}
\begin{tabular}{lccccc}
\hline
Galaxy     & N$\mathrm{_{PNe}}$& N$\mathrm{_{PNe}}$& K-S    &    D     & Number  \\
                 &   Total                        &          used                 &           &           & of modes\\
\hline
LMC$^b$  & 164                           &  158   & 0.999  & 0.037 &     2       \\
SMC$^c$  &  59                            &  55      & 0.970  & 0.097 &     2        \\
NGC\,682 & 23                              &  23     & 0.999 & 0.087 & 2\\
NGC\,3109 & 20                            &  19     & 0.993 & 0.136 & 1\\
NGC\,3109$\dag$ & 20                 &  19     & 0.888 & 0.206 & 2\\
NGC\,300   & 100      &  95     & 0.999 & 0.052 & 1\\
M\,33          &152      &  144    & 0.995 & 0.047 & 2\\
M\,31         & 298 &  288    & 0.879 & 0.048 & 2\\
NGC\,205   & 35       &  32     & 0.795 & 0.156 & 1\\
NGC4697   & 535     &  420   &  0.95  & 0.14   &  1 \\
\hline \hline
\end{tabular}
\end{table}

\section{Are two modes reasonable?}

In the majority of the galaxies in our sample, in particular irregulars and spirals, we have
shown that the PNLF can be fitted  with 2 modes, 
equation~(\ref{eq:2cumulative}). 
At the same time many studies have shown that the value 
of the  absolute magnitude of the brightest planetary nebula is fairly
uniform in a large sample of galaxies (C89). This could be explained
if PNe are initially formed at a similar brightness $M^*$.   The maximum brightness is attained for PNe with central stars having evolved from stars with initial masses around 2.5 solar masses (see figure. 10 in Marigo et al. 2004). If galaxies with old stellar populations have the same $M^*$ as star forming galaxies, this means that their brightest PNe should have progenitors of equal masses. This is against canonical stellar evolution, unless galaxies with old stellar populations have all experienced a recent star formation burst and there is no evidence for this. One way  out of this problem is the scenario proposed by Ciardullo et al. (2005) in which the brightest PNe arise from coalescence of two intermediate-mass stars.

The expansion of a PN will produce a drop in density, thus 
reducing its brightness. The decrease in brightness, in the case of ionisation bounded PNe
is also due to the decrease in the luminosity of the central star when it enters the
 white dwarf cooling track. As time proceeds, new bright PNe will be produced in the
galaxy (from less massive stars) to replace those that have become fainter,
and eventually  the entire PNe distribution function will be shifting to larger magnitudes,
(Marigo et al. 2004). 

In order to study the two-population PNLF  one can consider the lifetime 
of a planetary nebula $\sim 10^3$-- $10^5$~yr, which is small  compared 
to the star progenitor lifetime. Thus, to some extent, the PNe that we
observe are being produced by stars dying at the present time.

Let us consider a stellar initial mass function defined by Kroupa et al. (1993),

\begin{equation}
\phi(m)=\phi_0 m^{-\alpha}
\label{eq:imf}
\end{equation}

where $\phi_0$ is a normalization constant,

\begin{equation}
\alpha=\left \{
  \begin{array}{lc}
  1.3;   &  M_{l} < m < 0.5~{\rm M}_\odot\\  
  2.3;   &  0.5 < m < 1.0~{\rm M}_\odot\\  
  2.8;   &  1.0 < m < M_{u}\\  
\end{array}
\right.
\end{equation}

where  $M_l=0.1$~M$_\odot$ and $M_u$=100~M$_\odot$ are the lower and 
upper limits of the initial mass function.
The number of stars, or PN observed nowadays, can be
related to the star formation history $\psi(t)$ (mass per unit time
turned into stars at a given time $t$) as

\begin{equation}
\frac{dN_{PNe}}{dt} \propto \phi(m)\psi(t-\tau),
\label{eq:nps}
\end{equation} 

where $\tau$ is the time elapsed since the last busrt.

Therefore, the total number of PNe for two populations
(labeled $1$ and $2$) can be estimated by, 

\begin{eqnarray}
N_1&=&\psi(t-t_1)\phi(m) \Delta t_1,\label{eq:N0}\\
N_2&=&\psi(t-t_2)\phi(m) \Delta t_2,\label{eq:N1}
\end{eqnarray}
where $t_1$ and $t_2$ are the time since the formation of each
stellar population, assumed to have happened in  bursts of width
$\Delta t$. These equations also assume that populations 1 and 2
have PNe with the same lifetimes.

Since PNe are formed by stars within a range of
masses from $\sim$1 to  8~M$_\odot$, we should restrict the
initial mass function to this range

\begin{equation}
\label{eq:imf}
\phi(m)=\int^{8 \rm{M}_\odot}_{1 \rm{M}_\odot} \phi_0 m^{-\alpha} dm.
\end{equation}

Taking the lifetime of the stars to be a function of their mass,
 $\tau(m)$, we can estimate the age of the stars dying now for each
population,
 
\begin{eqnarray}
\tau(m)_1&=& t_f-t_1 ,\nonumber\\
\tau(m)_2&=& t_f-t_2,
\label{eq:tau}
\end{eqnarray}

where $t_f$ is the age of the host galaxy. The main sequence lifetime
$\tau_{MS}(m)$ can be obtained, for instance, from stellar evolution models
(e.g. Maeder \& Meynet 1989).
One can, however, invert the equation  and obtain the mass of the stars
that  are progenitors of the PNe seen at the present time for each population. 
The ratio between the number of PNe from the first and
the second population is given by (equations~\ref{eq:N0} and \ref{eq:N1}), 
\begin{equation}
\label{eq:ratio}
\frac{N_2}{N_1}=\frac{\psi(t-t_2)}{\psi(t-t_1)}\frac{\Delta t_2}{\Delta t_1}\left(\frac{m_1}{m_2}\right)^{-1.35}
\end{equation}

Carigi et al. (2006) , based on the spectrophotometric study of Wyder (2001 and 2003), obtained the star fomation history of NGC\,6822 
and found evidence for two separate events of star formation. From their results we can estimate the $N_1/N_2$ ratio of
the two stellar populations. 
The NGC\,6822 estimated age is $\sim$13.5 Gyr, and the 
peak of each starburst of this galaxy (see figure 4 of Carigi et al. 2006), occurred at 
$t_1=8$ Gyr and $t_2=11.5$~Gyr.
Using these values in equation~\ref{eq:tau}, we obtain $\tau(m)_1=5.5$~Gyr and 
$\tau(m)_2=2$~Gyr. From the models of 
Maeder \& Meynet (1989) we can estimate the mass of the stars dying at
the present time to be $m_1\sim 1.45$~M$_\odot$ and 
$m_2\sim 1.8$~M$_\odot$.

For the two bursts of star formation derived
by Carigi et al. (2006) we then have:

\begin{itemize}
\item burst 1: $t_1=5$ Gyr ago. The stars from this burst which are
  producing PNe today have a mass of $m_1\approx 1.4$M$_\odot$,
\item burst 2: $t_2=0.74$Gyr ago. The stars from this burst which are
  producing PNe today have a mass of $m_2\approx 2.5$M$_\odot$.
\end{itemize}

\begin{figure}
\centering
\includegraphics[width=\columnwidth]{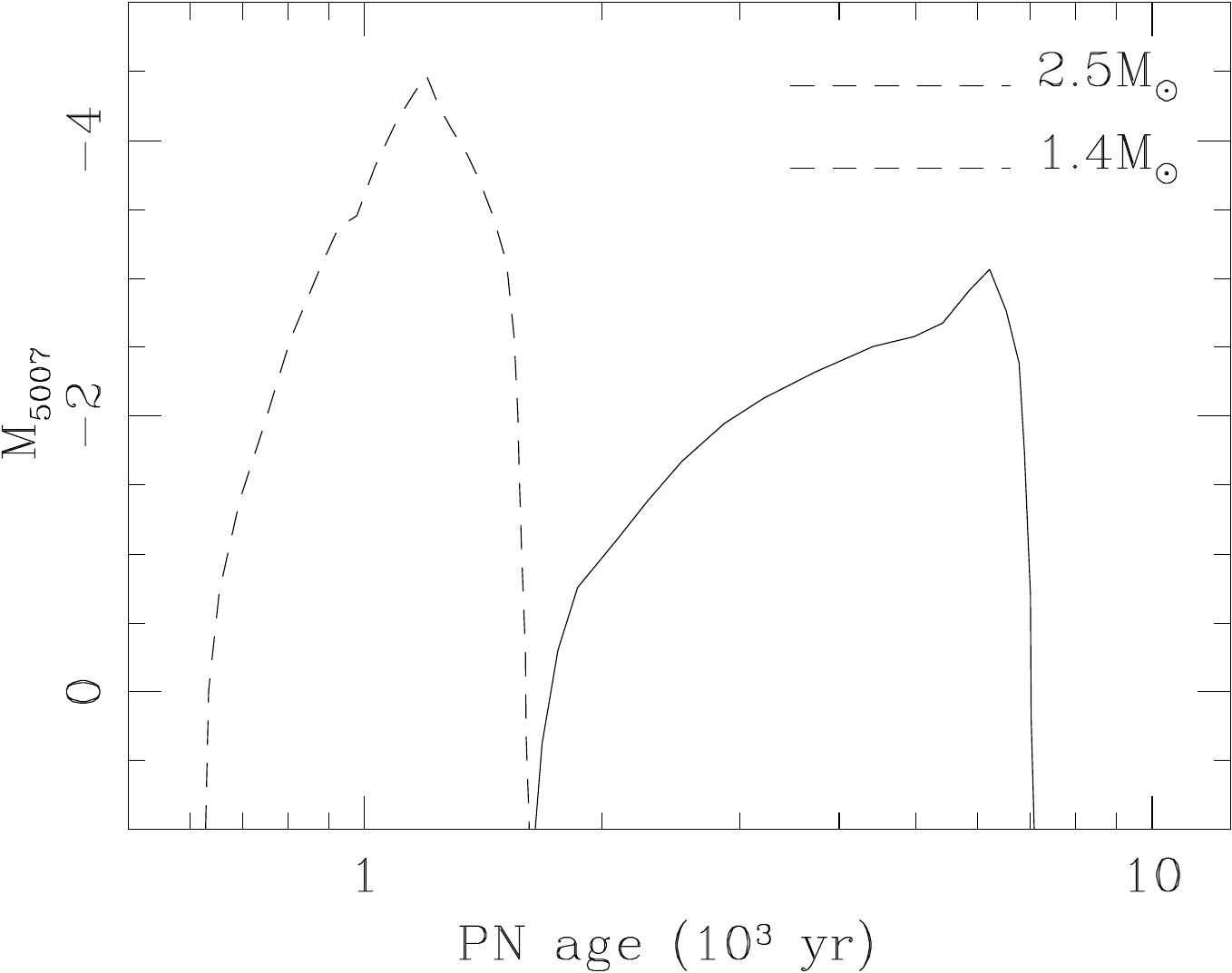}
\caption{$M_{5007}$ (absolute magnitude of the  5007 line) versus PN
age for two progenitor masses, 2.5 (dashed line) and 1.4 (solid line) M$_\odot$
(taken from figure 10 of Marigo et al. 2004).}
\label{f:mage}
\end{figure}

For each of the two bursts, we will have PNe corresponding to stars
of masses $M_1$ and $M_2$ (respectively) at all stages of their
evolution since the nebulae evolve and dissipate in timescales of
$\sim 10^4$yr, which are much smaller than the duration of the bursts.
In Figure~\ref{f:mage}, we show the [O III] magnitude $M_{5007}$ as a function of
evolutionary time $t_{ev}$ (from the beginning of the PNe phase) of the nebula
produced by stars of masses $M_1$ and $M_2$, taken from Figure 10 of
Marigo et al. (2004).

If we assume that the star formation rate was
uniform within the $\sim 10^4$yr total evolutionary time of the
observed PNe, from the M$_{5007}(t_{ev})$ curves of Figure~\ref{f:mage} we can
obtain the frequency distribution of the nebula as:

\begin{equation}
f(M_{5007})=A\,\left[\left(\frac{dt_{ev}}{d_{M_5007}}\right)_{t_a}+\left(\frac{dt_{ev}}{d_{M_5007}}\right)_{t_b}\right]\,,
\label{fm}
\end{equation}

where $t_a$ and $t_b$ are the two intercepts between the
$M_{5007}=const.$ line and one of the $M_{5007}(t_{ev})$ curves of
Figure~\ref{f:mage}. $A$ is a normalization constant chosen such that $\int
f\,dM=1$.

\begin{figure}
\centering
\includegraphics[width=\columnwidth]{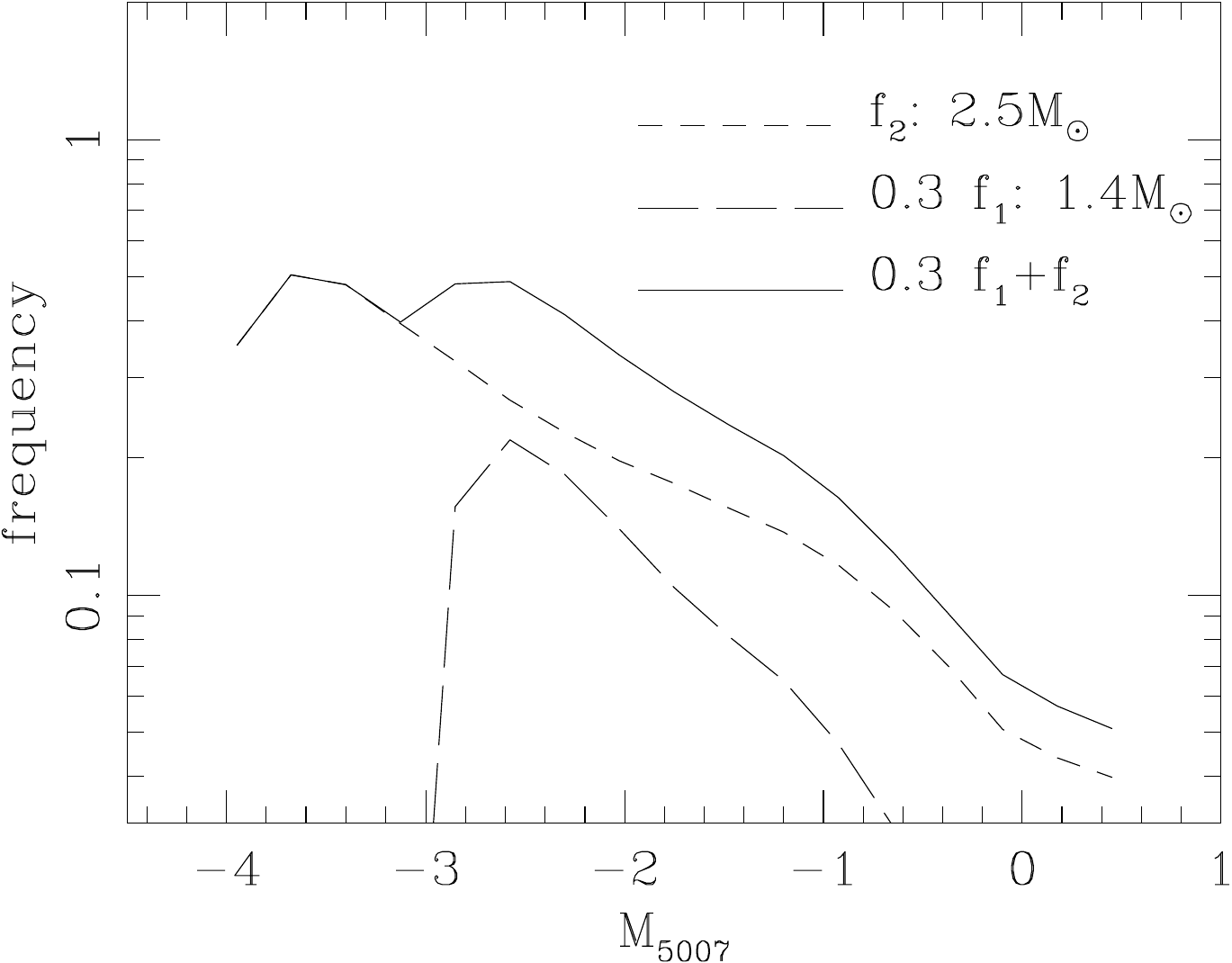}
\caption{Frequency rate for the PNe of the two burst versus
  $M_{50007}$. The short-dashed line shows the normalized frequency
  for the second burst, and the long-dashed line shows  
the normalized frequency (multiplied by $0.3$) for the first
burst. The solid line is the sum.
}
\label{f:freq}
\end{figure}

In this way, we have computed the $f(M_{5007})$ frequency rates for the
PNe of the two bursts ($f_1$ and $f_2$, respectively), which are shown
in Figure \ref{f:freq}. Also shown in this figure is the weighted sum
$f=0.3f_1+f_2$.

We see that this weighted sum of the PNe distributions estimated for
the two star formation bursts has two  modes  at about the same height,
with a horizontal separation of $\Delta M_{5007}\sim 1.2$. This distribution
function is similar to the distribution which we have
obtained from the PNe observed in NGC 6822  (see Figure {\ref{f:6822Can}}), which also has two peaks of similar heights,
separated by $\Delta m_{5007}\sim 1.4$.

Therefore, we find that this simplified model
 qualitatively reproduces the general morphology of the NGC 6822 PNe
 distribution function.
In order to obtain two peaks of similar heights (observed in the NGC
6822 distribution) it is necessary to have a second episode (creating the $\sim
2.5$M$_\odot$ stars giving rise to some of the present day PNe) which is
substantially more massive than the first star formation episode
(which created the $\sim 1.4$ M$_\odot$ stars). 
 This study could be extended to other galaxies which present a
  dip in the PNLF and a star formation with more than one burst
  (e.g. SMC, see No\"el et al. 2008).

\section{Summary}

We have proposed an extension of the planetary nebula luminosity
function of C89 and J90 to include two different stellar populations, and
we have applied it to a sample of PNe in 9 galaxies.

The most important parameter of the PNLF is $M^*$, the magnitude of the
brigthest PNe expected in the PNe population, which has been found to be
remarkably similar in many galaxies, and is therefore used as a distance
estimator.

The original C89 and J90 PNLF adjusted 2 parameters for one population:
$m^*$ and $N_{T}$, which are the magnitude of the brightest PN
and a normalisation constant, respectively.
 To obtain these parameters with a large sample of PNe one could fit
the functional form of the PNLF to a histogram of the sample. On the
contrary, if the sample is small, as it is in many cases, a maximum
likelihood method has to be used.
In any case the data has to be restricted to the brightest PNe due to
sample incompletness.

In our two-mode PNLF a first mode dominates the brightest part of the sample,
while the second mode becomes comparable, or dominant at fainter magnitudes.
We have introduced a parameter $m_{cut}$ for the first mode, a
sharp cutoff at the faintest magnitude. This is not included for the second
mode to have as few parameters as possible. For the second mode the 
cutoff is artificially placed where the sample becomes incomplete. Therefore
 we can in general fit two populations in a PNLF with 5 parameters, all of
which go into a genetic fitting algorithm.
 
To obtain the parameters in our PNLFs we used a genetic algorithm to
fit the cumulative luminosity function, which can be used with small
or large samples.
Since the two-mode PNLF considers the presence (or absence) of a dip
in the PNLF we cover a larger magnitude range than what is typically
used with a single mode PNLF.

In order to explore the PNLF for galaxies of different Hubble types,
we have selected 9 galaxies:  four irregulars (SMC, LMC, 
NGC\,6822, NGC3109), three spirals (NGC\,300, M31 and M33), the dwarf elliptical 
NGC\,205, and the elliptical NGC\,4697. 
We have fitted the two-mode PNLF to each of these galaxies and find a good
agreement with the $m^*$ estimated by other authors.
Our fitting procedure automatically finds a single mode fit (returns 
a $m_{1}\sim m^*_{2}$) where the data is more consistent with
a single mode.
All of the irregular galaxies in our sample, except one, are consistent with two modes.
The situation is less clear for spiral and elliptical galaxies, and the sample of galaxies used here is not
enough to see any trend of the number of modes with the Hubble type. However, we can see a little trend 
to the star forming galaxies to present a PNLE with 2 modes. 

Finally, comparing the results of our two-mode fit to the PNLF of  NGC\,6822
with the star formation history in this galaxy given by Carigi et al. (2006), we have shown that the two
modes (two bursts of star formation) reproduce a combined frequency of
PNe with two peaks separated by $\sim 1.2$ magnitudes, which is consistent 
with the sample of PNe observed in NGC\,6822.


\begin{acknowledgements} We acknowledge support from CONACyT grant
167625 and DGAPA-UNAM grants
IN105312 and IN106212. L. H-M \& M. P.  acknowledge the financial support from 
PAPIIT grant IN109614 (DGAPA-UNAM). L. H-M want to appreciate the hospitality of IA-UNAM.
\end{acknowledgements}

\end{document}